# Grain Boundary Motion Exhibits the Dynamics of Glass-Forming Liquids

*Hao Zhang*

Department of Chemical and Materials Engineering,
University of Alberta, AB T6G 2G6 Canada

*David J. Srolovitz*

Department of Physics, Yeshiva College, Yeshiva University, NY 10033 USA

*Jack F. Douglas\**

Polymers Division, NIST, Gaithersburg, Maryland, 20899 USA

*James A. Warren*

Metallurgy Division, NIST, Gaithersburg, Maryland, 20899 USA

*Polycrystalline materials can be viewed as composites of crystalline particles or 'grains' separated from one another by thin 'amorphous' grain boundary (GB) regions. While GB's have been exhaustively investigated at low temperatures, where these regions resolve into complex ordered structures accessible to measurement, much less is known about them at higher temperatures where the GB can exhibit significant mobility, structural disorder, and where experimental characterization methods are limited. The time and spatial scales accessible to molecular dynamics (MD) simulation make this method appropriate for investigating both the dynamical and structural properties of grain boundaries at elevated temperatures. In the present study, we use MD simulations to determine how the GB dynamics changes with temperature and applied stress. It has long been hypothesized that GB's have features in common with glass-forming liquids based on the processing characteristics of polycrystalline materials. We find remarkable support for this suggestion, as evidenced by string-like collective motion, transient caging of atom motion, and non-Arrhenius (Vogel-Fulcher) temperature dependence of GB mobility. Evidently, the frustration caused by the inability of atoms in GB region to simultaneously order with respect to competing grains is responsible for this striking similarity. The paradigm that grains in a polycrystalline material are encapsulated by a 'frustrated fluid' provides a powerful conceptual model of polycrystalline materials, pointing the way to improved control over their material properties.*

Most technologically important materials are polycrystalline in nature (1) and it is generally appreciated that the grain boundaries (GB) of these materials, the interfacial region separating the crystalline grains (see Fig. 1a), can have pronounced effects on the properties of this broad class of materials (2). In particular, the dynamical properties of GB, such as the rate of GB migration, play an important role in the plastic deformation and evolution of microstructure that occurs both during material processing and service (3). In inorganic crystalline materials, the GB's are often as wide as several nanometers (in metals the GB widths are more typically in the 0.2 nm to 1 nm range). When the grain size is on the nano-scale, the fraction of the material in these GB regions becomes appreciable [in excess of 10 % for polycrystalline materials with a grain size of $\approx$10 nm (4)] and the GB can then come to dominate material properties (5).

Equilibrium GB crystallography can be minimally specified by five variables: three to specify the relative orientation of one grain with respect to the other and two to indicate the boundary plane (inclination) (3) (Note the distinct orientations of grains indicated Fig. 1b; a *misorientation angle* describes a rotation of one grain that will cause it to coincide in orientation with the other.).While these variables determine the basic GB crystallography, they do not fix the atomic structure within the GB. This structure has variously been described in terms of dislocations (6), structural units (7), a coincidence site lattice (3), etc. Evidently, when the misorientation between two adjacent crystals obtains a certain angle, there will be common atomic positions for both crystals, and this defines the so-called coincidence site lattice (CSL) (3). Such lattices are conventionally characterized by a parameter $\Sigma$, defined as the ratio of the volume of an elemental cell in CSL to the volume of a lattice cell of the crystal lattice describing the individual grains.

Highly symmetric grain boundaries (having relatively small Σ) often exhibit special properties such as lower GB energy, higher GB mobility and a lower activation energy for migration (3). All of these descriptions of GB structure have their limitations at elevated temperatures (in relation to the melting point) where the GB's exhibit appreciable disorder.

The atomic positions adopted within the GB region represent a compromise between the perfect crystalline order of the adjacent grains, while maintaining the large scale GB crystallography. This competition leads to "packing frustration," which manifests itself in a relative disorder of the local atomic packing (hence the term 'amorphous' applies). Frustration in local ordering is also characteristic of glass-forming fluids, where molecular ordering tends to be of a relatively short-range in nature (8). This simple physical observation leads us to expect similarities between the molecular dynamics of glass-forming fluids and GB in the temperature ($T$) dependence of relaxation processes (e.g., GB migration, transport processes in the GB region). Moreover, the strong correspondence between the dynamics of glass-forming liquids and GB molecular dynamics that we discuss below leads us to expect GB migration to be sensitive to impurities, geometrical confinement and the mode of applied strain (compression versus tension), since factors that affect molecular packing have been shown to significantly influence the dynamics of glass-forming liquids (9-11).

The idea that the atomic dynamics in the GB region has an intrinsically 'glassy' nature has been proposed before, but the lack of a quantification of these prior claims has made these assertions open to challenge. Nearly 100 years ago, Rosenhain and Ewen (12) suggested that metal grains in cast iron were "cemented" together by a thin layer of 'amorphous' (i.e., non-crystalline) material "identical with or at least closely analogous to the condition of a greatly undercooled liquid." While this model, which suggested that iron and certain steels are composed of crystal grains coated with a "cement-like" layer of glassy metal, was able to rationalize a wide range of processing characteristics of ferritic materials (12), it was not possible at the time to validate this model through direct observation or simulation. Despite its early successes, this model of polycrystalline metals fell into disfavor and was subsequently ignored. Sixty years later, Ashby (13) used a model system involving layers of macroscopic bubbles ('bubble rafts') to "simulate" the atomic motion that occurs in GB migration and concluded that the atomic dynamics were consistent with the GB having an intrinsically glassy nature. Again, the inability to directly test this hypothesis in a real material limited its acceptance. Recent MD simulations of polycrystalline metals and silicon by Wolf and co-worker (14-16) suggest that the disordered structure within GB can be well described by atomic radial distribution functions characteristic of glass-forming liquids. Interestingly, a more recent MD simulation of atomic self-diffusion in the GB region again suggested a diffusion mechanism reminiscent of bulk metallic glasses (17).

In this paper, we address a central question in the high-temperature dynamical behavior of polycrystalline materials: "Do the transport properties of grain boundaries under relatively high temperature conditions have the characteristics of glass-forming liquids?" To answer this question, we perform a series of MD simulations of the migration of a high-angle general tilt boundary at elevated temperatures and compare the behavior of the GB migration to dynamical properties of classical glass-forming liquids. In particular, we simulate the migration of an asymmetric [010] general tilt boundary (misorientation angle $\theta$ = 40.23°; see Fig. 1b) in a pure Ni bicrystal (a model system including only two crystal grains and an interface between them) at temperatures ranging from 0.48 $T_m$ to 0.86 $T_m$, where the bulk melting temperature equals $T_m$ =1624 K (18). The interactions between Ni atoms were described using an embedded-atom method (19, 20) potential and the temperature was controlled using a Hoover-Holian thermostat (21). GB migration was driven by the difference in the stored elastic energy in the two crystals (the crystals are elastically anisotropic) (22, 23). The total number of atoms in the system was 22,630. The simulation cell was chosen such that the [010] tilt axis is parallel to the Y-direction, the misorientation angle between the two [100] axes of each grain is $\theta$ = 40.23° where the GB plane is perpendicular to the Z-direction (see Fig. 1b) in the visualization. Compressive strains were applied in the X- and Y- directions, while the (+Z) top and (-Z) bottom surfaces of the bicrystal are free (22, 23).

Both atomistic simulations and experiments demonstrate that string-like cooperative atom motion is a common feature of the dynamics of glass-forming liquids (24-32). In order to examine whether a similar dynamics occurs in the GB regions of polycrystalline materials, we apply the same techniques originally developed to identify such collective behavior in glass-forming liquids (24) to MD simulations of GB migration. First, we identify the 'mobile' atoms in our system, i.e., those that move a distance in a time $\Delta t$ that is larger than the typical amplitude of an atomic vibration, but smaller than the second nearest-neighbor atomic distance. Then, the mobile atoms $i$ and $j$ are considered to be within a displacement string if they remain nearby one another as they move (see Fig. 1c). Further details of this procedure are provided in (18) and (10).

The scale of cooperative motion in glass-forming liquids can be described by determining the average 'string length',

$$\bar{n}(\Delta t) = \sum_{n=2}^{\infty} nP(n,\Delta t), \qquad (1)$$

where $P(n,\Delta t)$ is the probability of finding a string of length $n$ in time interval $\Delta t$. 'Strings' are contiguous group of mobile atoms that remain in close proximity while they are moving. Previous work (9, 23, 33-35) has established that the average length of these strings in glass-forming liquids grows in parallel to the apparent activation of structural relaxation, in accord with the Adams and Gibbs theory of relaxation in glass-forming liquids (36) and the identification (24) of the strings with the cooperatively rearranging regions of their theory. Strings are thus of interest since they are correlated with the relative strength of the $T$-dependence of transport properties (see below). Below we show how changes in the GB mobility due to compressive and tensile applied strains, which were previously puzzling to us, can be understood from the changes in the string properties under these deformation histories.

**Results and Discussion**

Figure 2 shows the temporal evolution of the average boundary position at eight different values of $T$. The position of the boundary was determined based upon the atomic coordination numbers (see caption of Fig. 1b). The boundary velocity $v$ was simply obtained from the average slope of the boundary position versus time where ultimate displacements from 1 nm to 5 nm were considered, depending on $T$. In the classical theory of GB migration (37), the $T$ dependence of the boundary mobility (ratio of the velocity $v$ to the driving force $\Delta p$) obeys an Arrhenius temperature dependence, i.e.

$M = M_0 \exp(-Q/(k_B T))$, where $M_0$ is a constant and $Q$ is the activation energy for boundary migration. The inset in Fig. 2 (open circles) shows that this expectation is not satisfied. Instead, the fit to the GB mobility data to the Vogel-Fulcher equation (38),

$$M = M_{VF} \exp\left(-\frac{Q_{VF}}{k_B(T-T_0)}\right), \quad (2)$$

where $M_{VF}$ and $Q_{VF}$ are constants, is significantly better than the Arrhenius relation. This behavior is a characteristic, even a defining property, of glass-forming liquids (38). In this context, the temperature $T_0$ at which the mobility formally extrapolates to zero according is termed the Vogel-Fulcher (VF) temperature (38) and a best-fit to the data in Fig. 2 gives $T_0 = (509 \pm 18)$ K (error estimate is based on a 95 % confidence interval throughout this paper). We conclude that the $T$-dependence of the GB mobility obeys the same relationship as relaxation in glass-forming liquids, providing support for the physical picture of polycrystalline materials described in the introduction. We next examine the fitted parameters in Eq. (2) to further test the consistency between GB dynamics and glass-forming liquids.

First, we note that the ratio of the melting temperature $T_m$ to $T_0$ that we find is typical value of glass-forming liquids. For example, in metallic glasses the ratio of the melting (eutectic) temperature to $T_0$ is estimated to be $T_m / T_0 \approx 2.8$ (39), which is reasonably close to the corresponding ratio 1624 K / 509 K = 3.2 for GB migration. The temperature $T_m$ basically defines the onset and $T_0$ the end of the glass transformation process. The 'middle' of this transitional range is described by the so-called 'mode-coupling' temperature $T_c$ and the glass-transition temperature $T_g$, which are estimated below since these quantities are conveniently addressed in our simulations by a direct examination of changes in the molecular dynamics upon cooling. We next consider the commonality between the dynamics of GB motion and glass-forming liquids at the level of molecular dynamics and determine other characteristic temperatures in relation to the molecular dynamics of the GB region.

As discussed above, string-like cooperative atomic motion is prevalent in all glass-forming liquids examined to date [including water, polymer fluids, metallic glass-forming liquids, concentrated colloidal suspensions and even strongly driven granular fluids (9, 24, 28, 40, 41)]. In these fluids, the strong reduction in the atomic mobility and enormous changes in the rate of structural relaxation are associated with the growth of string-like correlated motion upon cooling. To this end, we now consider the nature of the atomic motion that occurs during GB migration. Figure 1c shows a typical displacement string that appears in the GB region during our simulation. In this figure, the initial atom positions are shown in yellow ($t = 0$) and their positions at time $\Delta t$ later are shown in blue (displacements are shown using arrows; see Figs. 2c and 2d)). Note that when one atom moves, another moves into its previous location, leading to a cooperative displacement-string. Moreover, the string-like cooperative motion in this GB occurs predominately along the direction parallel to the tilt axis (See Fig. 1d). In the tilt GB, particle displacement also tends to be fast in the tilt axis direction (42).

Atomistic simulations of glass-forming liquids suggest that the distribution of string lengths $P(n)$ is an approximately exponential function of $n$,

$$P(n) \sim \exp(-n/\langle n \rangle). \tag{3}$$

Figure 3 shows the distribution of string lengths at $\Delta t = t^*$, where the string length $\bar{n}(\Delta t)$ exhibits a maximum during GB migration. Interestingly, the distributions of string lengths in glass-forming liquids and in the GB are essentially the same (9). The average string length <$n$> depends strongly on $T$ and GB type. The magnitude of <$n$> is comparable to those found in glass-forming liquids for the corresponding $T$ range (see below.) We see that <$n$> increases upon cooling (see inset to Fig. 3) and that <$n$> is smaller for the high symmetry Σ5 GB than the low symmetry general GB. This suggests that atoms in the high symmetry GB are less frustrated than in the low symmetry case.

The observations that the $T$-dependence of the GB mobility is well described by the Vogel-Fulcher relation and the form of the string length distribution confirm the similarities between GB and glass-forming liquid dynamics. The physical origin of the non-Arrhenius $T$-dependence of the GB mobility is related to the cooperative (string-like) atomic motion within GB, as previously discussed for the case of glass-forming liquids (9, 23, 33-35). Since the scale of cooperative motion (string length) decreases with increasing $T$, higher $T$ implies lower energetic barriers for collective atomic transport. We next turn to a consideration of other quantitative aspects of the phenomenology of glass-forming liquids that have significance for understanding the transport properties of polycrystalline materials.

The dynamics of glass-forming liquids is characterized by a number of temperatures, including $T_A$ (demarking the onset of the cooperative atomic motion), a temperature $T_c$ (separating a high-$T$ regime of glass-formation from a low-$T$ regime where viscolelastic effects become prevalent), the glass-transition temperature $T_g$ (below which aging and other non-equilibrium behavior is overtly exhibited), and finally $T_0$ (characterizing the 'end' of the broad glass transformation process) (33-35, 43, 44). To further our analysis, it is natural to consider the analogous characteristic temperatures for the GB molecular dynamics. $T_A$ is experimentally defined as the $T$ at which the Arrhenius temperature dependence of structural relaxation no longer holds. Based on the entropy theory of glass-formation (44), the apparent activation energy $E_a(T)$ below $T_A$ follows a universal quadratic temperature dependence, $E_a(T)/E_a(T \to T_A) \approx 1 + C_0(T - T_A)^2$, where $C_0$ is a constant. Our GB migration data fits this relation well near $T_A$ (950K < $T$ < 1400K) and we estimate $T_A$ to equal $T_A$ = 1546 K. To estimate other characteristic temperatures, we performed a series of simulations to determine the mean square displacement of atoms <$r^2$> within the GB, following procedure described before by Starr *et al.* for glass-forming liquids (45).

Following Starr *et al.*, we define the Debye-Waller factor (DWF) as the mean square atomic displacement <$r^2$> after a particular decorrelation time $t_0$ characterizing the crossover from ballistic to caged atom motion (45). In glass-forming liquids, <$r^2$> exhibits a plateau after the time $t_0$ that persists up to the structural relaxation time of the fluid $\tau$ (normally a time many orders of magnitude larger than $t_0$) and the magnitude of this plateau defines the 'cage' in which particles are transiently localized by their neighbors. Measurements of <$r^2$> performed over a wide range of times (or frequencies) determine essentially the same value of <$r^2$> ≈ <$r^2(t_0)$> ≡ <$u^2$> for $t_0 < t < \tau$, explaining why dynamic neutron and x-ray scattering measurements that probe timescales on the order of $10^{-9}$ s are of relevance to the physics of glass-formation. Figure 4 shows <$u^2$> for the GB atomic motion as a function of $T$ (inset shows original <$r^2$> data) where <$u^2$> is indicated in units of the potential range parameter $\sigma$, a quantity that is roughly equal the inter-atomic distance. $T_c$ is estimated according to the standard recipe for glass-forming liquids (46). For 1050 K ≤ $T$ ≤ 1150 K, we performed a series of simulations to calculate the GB self-intermediate scattering function (47), $F_s(\mathbf{q},t) = \langle \exp\{-i\mathbf{q}[r_i(t) - r_i(0)]\} \rangle$, which is simply the Fourier transform of the atomic displacement distribution function $G_s(\mathbf{r},t)$ (The Fourier transform variable $\mathbf{q}$ is often termed the scattering 'wave vector'.) The variation of $F_s(\mathbf{q},t)$ with temperature is characteristic of glass-forming liquids and an examination of this quantity offers a good opportunity to examine how much GB atomic motion resembles the dynamics of glass-forming liquids.

After exhibiting a plateau that is associated with the particle caging phenomenon illustrated in Fig. 4, $F_s(\mathbf{q},t)$ in glass-forming liquids normally exhibits a 'stretched exponential' variation, $F_s(q,t) \propto \exp[-(t/\tau)^\beta]$. Figure 5 shows that this behavior is also exhibited in the GB atomic dynamics. Moreover, the stretching exponent $\beta$ determined from our fits in Fig. 5 is insensitive to temperature ($\beta \approx 0.34$) where the secondary decay is prevalent and where the structural relaxation time $\tau$ can be well-described (see Fig. 5 inset) by an apparent power law, $\tau \propto (T - T_c)^\gamma$ in the $T$-range indicated. Both of these observations on GB dynamics are also typical for glass-forming liquids (47), where the fitted temperature $T_c$ is conventionally termed the 'mode-coupling temperature'. In particular, $T_c$ and $\gamma$ from our GB dynamics simulations are estimated to equal, $T_c = 923 \pm 23$ K with $\gamma = -2.58 \pm 0.4$. The value obtained for $\gamma$ is also typical of glass-forming liquids (47).

Note that the apparent power law temperature scaling of $\tau$ in glass-forming liquids is restricted to an intermediate $T$ range between $T_A$ and the 'critical temperature' $T_c$ and correspondingly our GB data is similarly restricted to this $T$ range. Mode-coupling theory is an idealized mean field theory of the dynamics of supercooled liquids and the divergence in $\tau$ predicted by this theory does not actually occur. Instead, $T_c$ prescribes a 'crossover temperature' separating the high-$T$ and low-$T$ regimes of glass-formation (44). $T_c$ is a thus useful and widely utilized characterization temperature of glass-forming liquids, despite its theoretical shortcomings.

To complete our comparison of the characteristic temperatures of glass-forming liquids with those of mobile GB regions, we must determine the low-$T$ characteristic temperatures, $T_g$ and $T_0$. However, equilibrium simulations of liquids are normally limited to $T > T_c$ because of the growing relaxation and equilibration times of cooled liquids. The same difficulty holds for studying GB motion. Thus, $T_g$ must be obtained through *extrapolation* of high-$T$ simulation data. With this difficulty in mind, we observe that $<u^2>$ in Fig.4 exhibits a linear $T$ dependence up to about 940 K, a temperature near $T_c$. Temperatures above $T_c$ then define a different regime of behavior for $<u^2>$ and other properties related to it. An extrapolation of the $<u^2>$ data in Fig. 4 to 0 indicates a $T$ close to the VF temperature $T_0$ that was determined from our grain mobility data in Fig. 2, as found before by Starr *et al*. for the structural relaxation time in a previous simulation study on glass-forming liquids (45). Thus, we find another striking correspondence between GB molecular dynamics and the dynamics of glass-forming liquids.

Physically, the glass-transition temperature $T_g$ corresponds to a condition in which particles become localized in space at essentially random positions through their strong interaction with surrounding particles and an (arguably non-equilibrium) amorphous solid state having a finite shear modulus emerges under this condition. A particle localization-delocalization also underlies crystallization and the Lindemann relation is known to provide a good rough indicator of the melting transition. Correspondingly, the same instability condition has been advocated generally for glass-forming liquids (48), which allows a direct estimation of $T_g$ (43, 44). Following Dudowicz *et al*. (43, 44), we *define $T_g$* by the Lindemann condition $<u^2>^{1/2} = 0.125$ and we estimate $T_g$ = 695 K, which is a typical magnitude for metallic glass-forming liquids (49). We have now defined all of the characteristic temperatures describing GB dynamics and we next compare to the corresponding temperature relationships of glass-forming liquids.

A precise comparison of the dynamics of GB and glass-forming liquids requires some discussion of a conventional jargon used to classify different types of glass-forming liquids. An Arrhenius temperature dependence of structural relaxation is normally taken as an 'ideal' behavior and the relative deviation from it is quantified by the abstract term, 'fragility'. Fluids in which this deviation is large are said to be 'fragile', based on the presumption that this phenomenon derives from a high susceptibility of the fluid's exploration of its potential energy landscape to the changes with temperature. This terminology is notably different from engineering terminology where fragile fluids are termed 'short' because there is little time to work with the material before it solidifies due to the relatively rapid change of mobility upon cooling.

It is established phenomenologically that the glass transition occurs over a narrower temperature range in fragile liquids so that *ratios* of the characteristic temperatures of glass-formation provide a convenient, objective and model independent measure of fragility that can be used to classify different categories (43, 44) of glass-forming liquids. This classification should then allow a better understanding of how the GB dynamics relates to the dynamics of glass-forming liquids.

First, we see that the relatively large temperature ratio $T_c / T_g \approx 1.33$ for the GB dynamics corresponds to a 'strong' glass-forming liquid in comparison to polymeric glass-formers, but larger values of this ratio are found in network forming glasses such as $B_2O_3$ and $SiO_2$ which are at the 'strong' end of fragility spectrum (meaning simply that that relaxation in these systems is nearly Arrhenius at all temperatures). Metallic glass-forming liquids are thus *intermediate fragility* glass-formers. Some polymers having a simple monomer structure, and a flexible chain backbone and sidegroups can also be characterized as having an intermediate fragility (strong by the standards of polymers alone) and this class of fluids is a natural candidate for comparison with the GB dynamics. Table I summarizes the *T* ratios found in our GB dynamics simulations and the corresponding analytically predicted ratios (43, 44) for low molecular mass polymeric glass-forming liquids having a similar *intermediate fragility* [FF polymer model of refs. (43) and (44)]. This conclusion is independently confirmed through a consideration of the GB mobility data in Fig. 2 which yields a 'fragility parameter' $D = Q_{VF} / k_B T_0$ estimate of $D = 3.59 \pm 0.3$ where $Q_{VF}$ and $T_0$ are obtained by fitting the GB mobility data to Eq. (2). Theoretical estimates for *D* range from about 5 to 3 for strong and fragile high molecular mass polymeric glass-forming liquids, respectively (44). Again an intermediate fragility is indicated for the GB mobility data. In Table I, we also include a comparison to recent simulation estimates of the corresponding temperature ratios for a relatively strong glass-forming polymer liquid, rendered a strong glass-former through the addition of a molecular additive (10). We thus obtain yet another striking confirmation of at least a qualitative commonality between the dynamics of GB and glass-forming fluids. This correspondence applies both to the phenomenology of macroscopic transport properties (e.g., GB migration rate) and the molecular dynamics on an atomic scale.

**Table I   Characteristic Temperature Ratios**

|  | GB Dynamics | Model Strong Polymer Glass-Forming Liquids | |
| --- | --- | --- | --- |
|  |  | Analytic Estimates (43, 44) | Molecular Dynamics (10) |
| $T_c / T_g$ | 1.3 | 1.4 | 1.4 |
| $T_c / T_0$ | 1.8 | 1.7 | 1.5 |
| $T_A / T_c$ | 1.7 | 1.6 | 1.4 |
| $T_A / T_g$ | 2.2 | 2.2 | 2.0 |
| $T_A / T_0$ | 3.0 | 2.6 | 2.2 |

Our new paradigm for understanding the GB dynamics emphasizes the importance of string-like cooperative motion in understanding the transport properties of both polycrystalline materials and glass-forming liquids, and we now apply this perspective to investigate a formerly puzzling phenomenon relating to GB migration under large deformation conditions. Since the application of strain can be expected to influence molecular packing, and thus the extent of packing frustration in the fluid, we can naturally expect these and other perturbations that influence molecular packing (e.g., hydrostatic pressure, molecular and nanoparticle additives, nanoconfinement) to affect the collective string dynamics. In previous work, we found that a variation of $T$ and the grain boundary type both influenced the average string size (18) so this sensitivity of the string size to thermodynamic conditions is established. We can also expect that varying the type of loading conditions, such as applying compressive strain, tensile strain or even a constant hydrostatic pressure, will influence the character of string formation because these modes of deformation naturally affect molecular packing. These resulting changes in string geometry should then be directly reflected in the GB dynamics, providing an interesting test for our framework.

In previous simulations, we showed that compressive versus tensile deformations led to appreciable changes in the *T*-dependence of GB mobility (22), where the GB migration velocity differed by a factor on the order $O(10)$ under the tensile and the compressive deformation at 800 K. Given that the driving forces under tensile and compressive strain are comparable in magnitude, this sensitivity of the migration velocity or mobility to the mode of strain is difficult to explain in terms of conventional GB migration theories. A reexamination of our former simulation results from our new perspective indicates that the average string length in the 2 % tensile strain case and 2 % compressive strain cases at $T = 800$ K are 1.63 and 2.13, respectively. If one assumes the apparent activation energy for GB migration can be scaled by the average string length, i.e., $Q \propto <n> E_0$, where $E_0$ is the activation energy near the melting point, the change by the average string length (and apparent activation energy) caused in tension and compression can readily account for the order of magnitude change in the GB mobility found in our former simulations. We suggest that the main origin of this shift in the scale of collective motion derives from a shift of $T_g$ with deformation, compressive deformation acting like an increase in the hydrostatic pressure, which generally increases $T_g$, and extensive deformation having the opposite effect. Temperature and pressure studies will be required to confirm this interpretation of the origin of the deformation-induced changes in GB mobility and the scale of cooperative GB atomic motion.

The addition of impurities and nanoscale confinement can also be expected to affect the cooperativity of atomic motions in strained polycrystalline materials, as recently shown in simulations of glass-forming liquids (9-11). Specifically, if the impurities help relieve packing frustration, then the scale of collective motion should be greatly attenuated (9) and the $T$-dependence of the GB mobility appreciably weakened (glass-formation becomes 'stronger'), while if the impurities disrupt molecular packing then the scale of collective motion should become amplified and the $T$-dependence of GB mobility should be greatly amplified. Large changes in GB mobility, and the resulting properties of polycrystalline materials, are then expected from the application of strains and the presence of impurities through the influence of these effects on the scale of collective motion in the GB region.

We conclude that the disordered atoms within the GB region of polycrystalline materials at elevated $T$, approaching the melting temperature, exhibit many features in common with glass-forming liquids. Highly cooperative atom motion, such as the string-like motion, can greatly affect the average rate of GB motion and presumably other transport properties of polycrystalline materials. This perspective on the dynamic properties of GB is expected to shed significant light on the mechanical properties of polycrystals. Indeed, we expect the properties of this essentially viscoelastic complex GB fluid enveloping the crystalline domains within polycrystalline materials to have a large impact on the plastic deformation of these materials; a perspective that has not been fully explored. This point of view is contrasted with recent work that attributes the viscoelastic effects of polycrystalline materials to simply the presence of solid crystalline grains within uncrystallized melts so that the system is conceived to be something like a granular material (50, 51). Within our paradigm, the 'frustrated' atoms within the GB region should exhibit a high sensitivity to impurities, pressure and geometrical confinement so that we can anticipate significant changes in the plastic deformation properties of polycrystalline materials arising from a modulation of the collective motion in the GB regions through these perturbations. This perspective offers the promise of an increased control of the properties of semi-crystalline materials based on an understanding of this phenomenon and better control of defect structures that arise in processing these materials. Although this conceptual view of polycrystalline materials was intuitively recognized by scientists and engineers involved in the fabrication of iron materials at the beginning of the last century (12), the present work puts this working model of the deformation properties of polycrystalline materials (50, 51) on a sound foundation through direct simulation.

**Acknowledgements:** The authors gratefully thank Robert Riggleman of the University of Wisconsin and Anneke Levelt Sengers of the NIST for helpful comments and questions about the work and acknowledge the support of the US Department of Energy/ Grant No. DE-FG02-99ER45797 and the National Institute of Standards and Technology. Useful discussions were also facilitated through coordination meetings sponsored by the DOE-BES Computational Materials Science Network program.

**Figure Legends**

Figure 1: Illustration of string-like cooperative atomic motion within a GB. (a) Schematic microstructure of polycrystalline metal. Different colors indicate the individual grain having different orientations and the black line segments represent GB. (b) Equilibrium boundary structure projected onto the X-Z plane for $\theta = 40.23°$ [010] general tilt boundary at $T = 900$ K (X, Y and Z axes are lab-fixed Cartesian coordinates, while [100], [010] and [001] refer to crystallographic axes). Upper and lower grains rotate relatively to each other by 40.23° along the common tilt axis [010]. The misorientation angle $\theta = 40.23°$ does not correspond to a special $\Sigma$ value ($\Sigma$ refers to the ratio of the volume of coincidence site lattice to the volume of crystal lattice if atomic sites that are coincident in the two crystal lattices on either side of the GB for certain misorientation.). The atoms are colored by their coordination numbers q (orange: q = 12; others: q < 12). The simulation cell was chosen to have grain boundary plane normal to be along Z-direction. (c) Representative string within GB plane. Yellow and blue spheres represent the atoms at an initial time $t = 0$ and a later time, $t^*$. (d) Snapshot of string-like cooperative motion within the grain boundary region at $T = 900$ K at $\Delta t = t^*$. The rectangular box illustrates the simulation cell in the X-Y plane.

Figure 2: Temporal evolution of the average boundary position at eight different $T$. Inset shows the logarithm of the boundary mobility as a function of $T$ [open circles versus $1/T$ (top axis) compare to an Arrhenius relationship while the filled circles consider a Vogel-Fulcher reduced variables description (bottom axis)].

Figure 3:    String length distribution function for Σ5 boundary and general GB at 800 K and 1400 K. The inset shows the average string length $<n>$ as a function of $T$ for the GB, illustrating the growth of the scale of collective motion upon cooling.

Figure 4    Debye-Waller factor as a function of $T$. The Vogel-Fulcher temperature $T_0$ was determined from fitting the $T$ dependence of boundary mobility data in Fig. 2. The temperature $T_A = 1546$ K approximates the onset of the supercooling regime was also determined from mobility data in Fig. 2. The crossover temperature $T_c$ is determined from the boundary relaxation time (See Fig. 5) and $T_g$ is estimated by the condition that $<u^2>^{1/2} \approx 0.125$, a Lindemann condition for glass formation (35). The Debye-Waller factor exhibits a linear $T$ dependence at low $T$ (harmonic localization) up to $T$ near $T_c$. At higher $T$ the dependence is super-linear. All the characteristic temperatures estimates correspond to a strong glass-forming liquid, as expected for metallic glasses.

Figure 5    The self-intermediate scattering function for GB particles in the $T$ range of 1050 K and 1150 K (defined in the text). Inset shows a power fit to the $T$ dependence of the structural relaxation time, $\tau$.

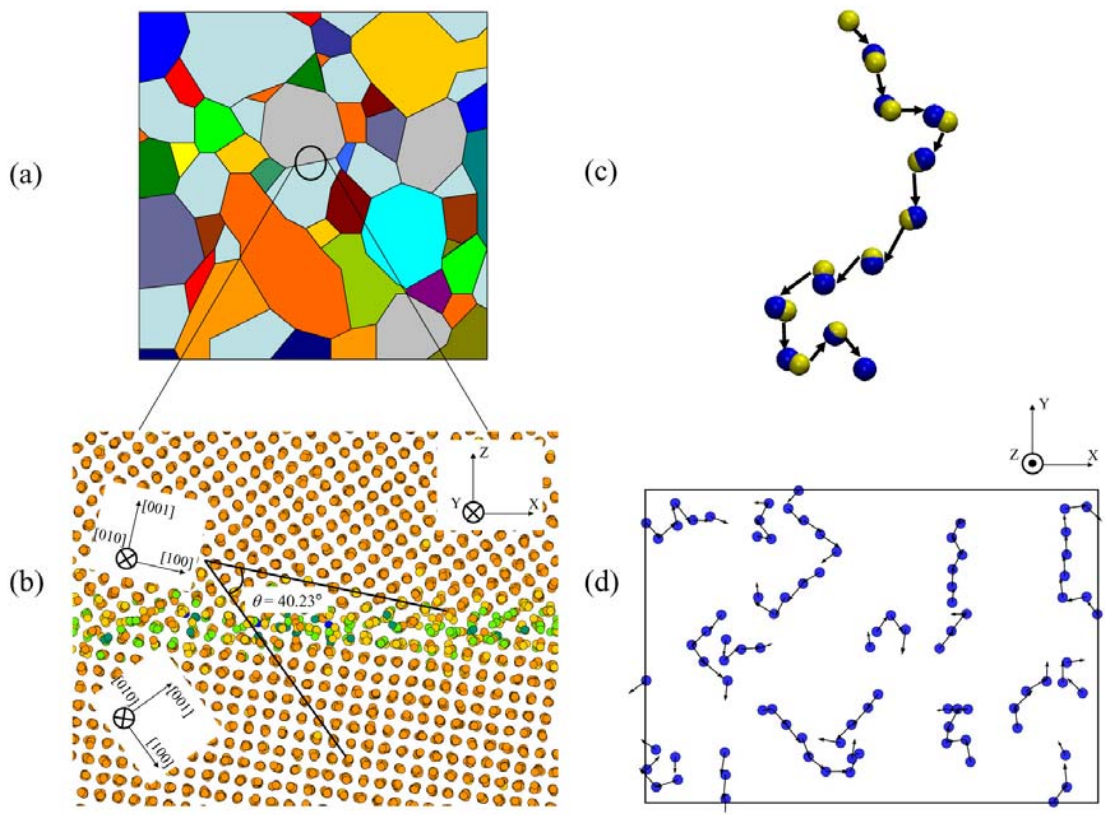

Figure 1

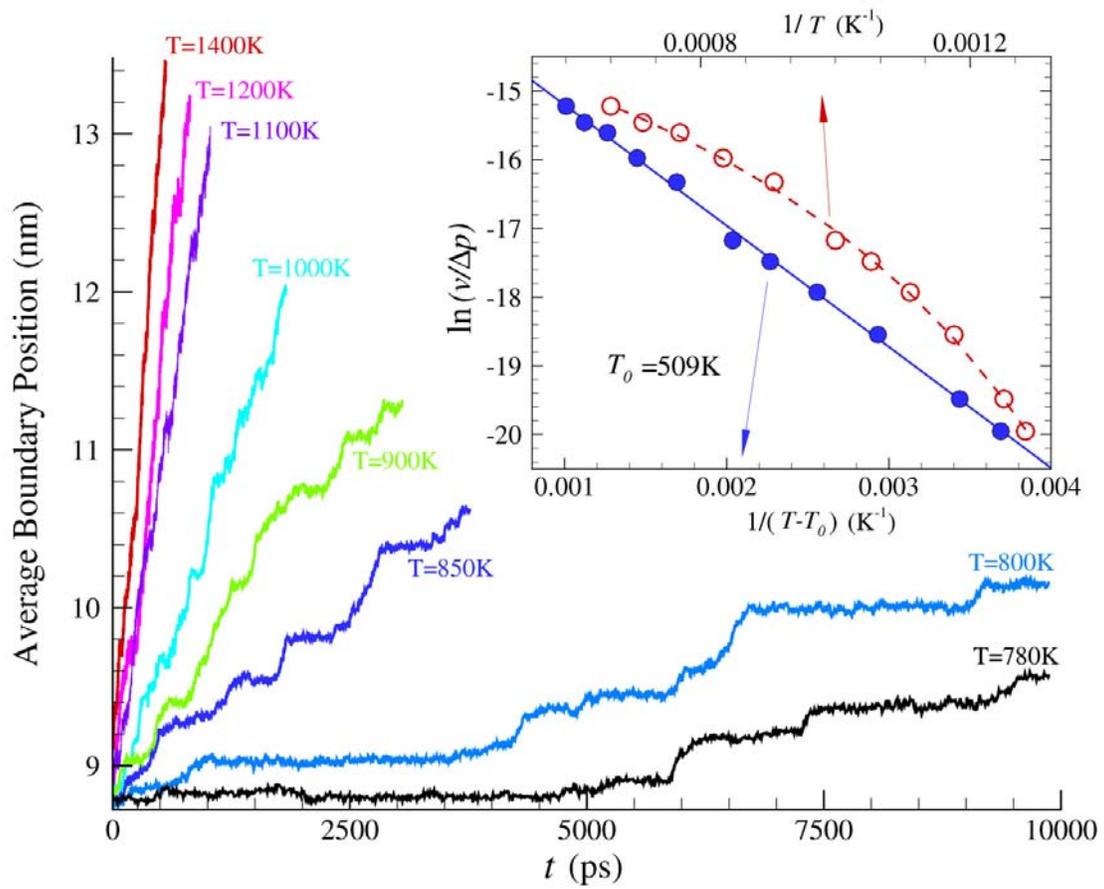

Figure 2

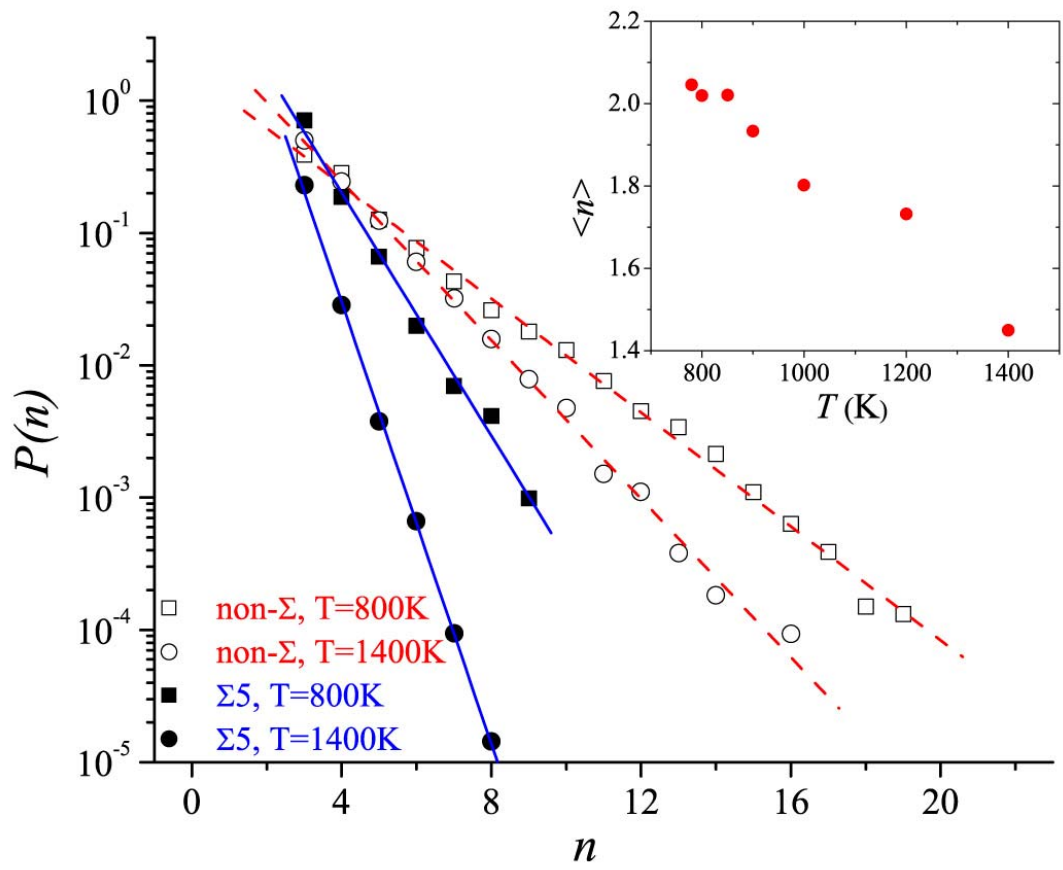

Figure 3

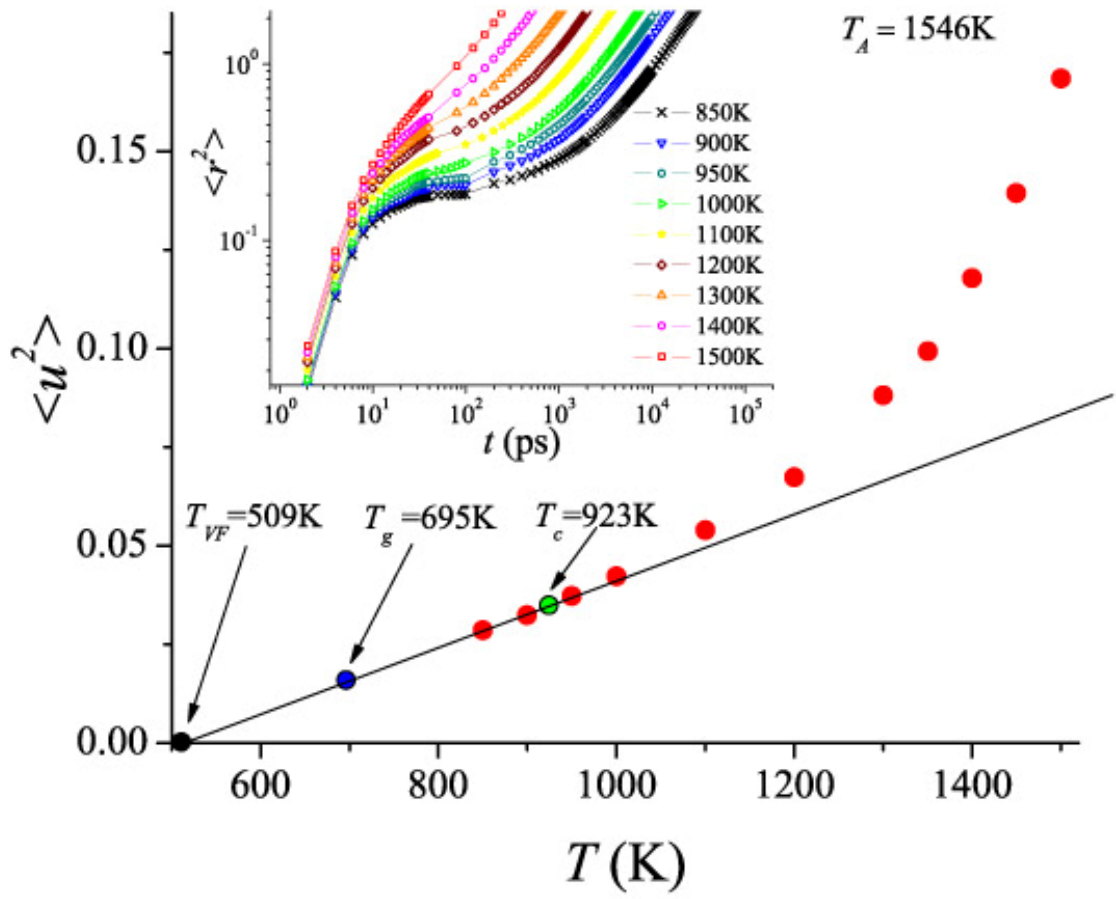

Figure 4

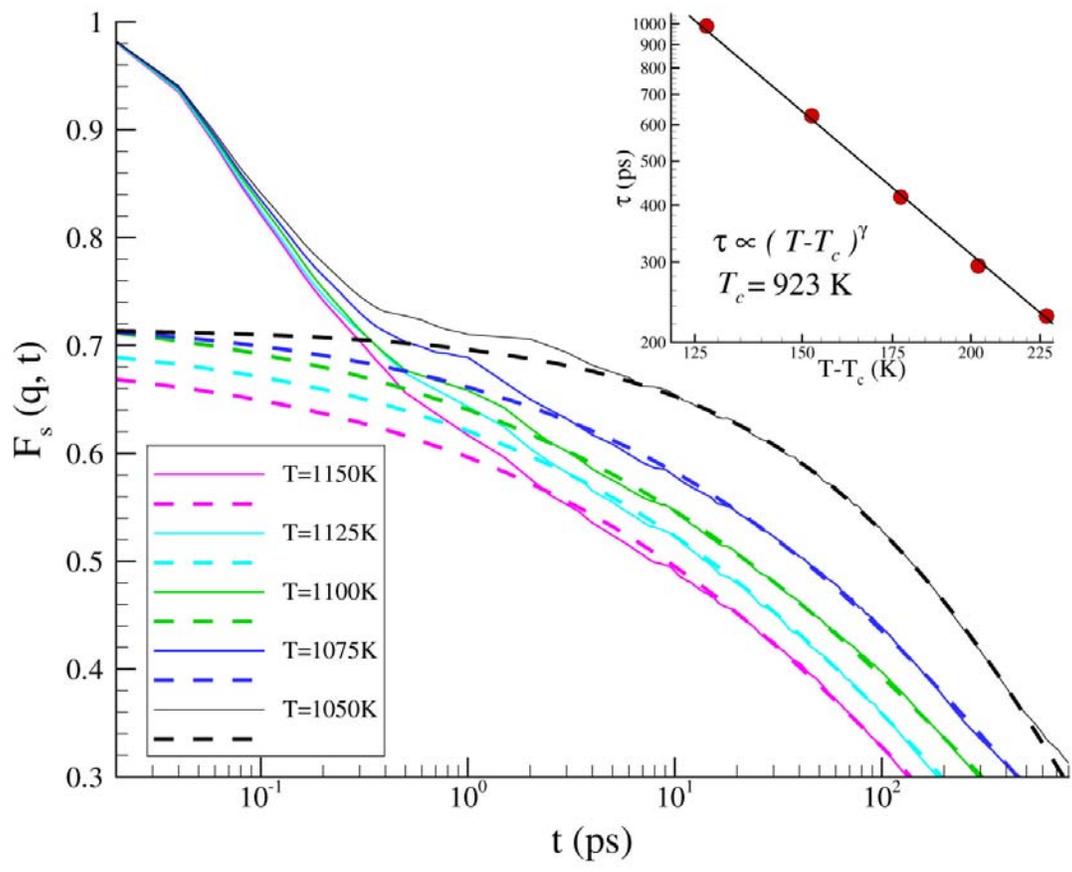

Figure 5